\documentclass[aps,twocolumn,superscriptaddress,showpacs,floats]{revtex4}

\usepackage{amsmath}
\usepackage{amsfonts}
\usepackage{amssymb}
\usepackage{euscript}
\usepackage{bm}
\usepackage[dvips]{graphicx}

\begin{document}

\title{On-site number statistics of ultracold lattice bosons}

\author{Barbara Capogrosso-Sansone}
 \affiliation{Department of Physics, University of
Massachusetts, Amherst, MA 01003}
\author{Evgeny Kozik}
 \affiliation{Department of Physics, University of
Massachusetts, Amherst, MA 01003}
\author{Nikolay Prokof'ev}
\affiliation{Department of Physics, University of Massachusetts,
Amherst, MA 01003}
 \affiliation{Russian Research Center
``Kurchatov Institute'', 123182 Moscow, Russia}
\author{Boris Svistunov}
\affiliation{Department of Physics, University of Massachusetts,
Amherst, MA 01003} \affiliation{Russian Research Center
``Kurchatov Institute'', 123182 Moscow, Russia}

\begin{abstract}
We study on-site occupation number fluctuations in a system of
interacting bosons in an optical lattice. The ground-state
distribution is obtained analytically in the limiting cases of
strong and weak interaction, and by means of exact Monte Carlo
simulations in the strongly correlated regime. As the interaction
is increased, the distribution evolves from Poissonian in the
non-interacting gas to a sharply peaked distribution in the
Mott-insulator (MI) regime. In the special case of large
occupation numbers, we demonstrate analytically and check
numerically that there exists a wide interval of interaction
strength, in which the on-site number fluctuations remain Gaussian
and are gradually squeezed until they are of order unity near the
superfluid (SF)-MI transition. Recently, the on-site number
statistics were studied experimentally in a wide range of lattice
potential depths [Phys. Rev. Lett. \textbf{96}, 090401 (2006)]. In
our simulations, we are able to directly reproduce experimental
conditions using temperature as the only free parameter.
Pronounced temperature dependence suggests that measurements of
on-site atom number fluctuations can be employed as a reliable
method of thermometry in both SF and MI regimes.
\end{abstract}


\maketitle

%%%%%%%%%%%%%%%%%%%%%%%%%%%%%%%%%%%
\section{Introduction}
\label{sec:intro}
%%%%%%%%%%%%%%%%%%%%%%%%%%%%%%%%%%%

Experiments on ultracold atoms trapped by an optical potential
\cite{Bloch_SFMI,Ospelkaus,Bloch_spin_dynamics,Ketterle_spatial,Bloch_SF3D},
offer a unique possibility to explore fundamental properties of
strongly correlated quantum many-body systems allowing virtually
unlimited control over the microscopic Hamiltonian parameters
(see, e.g., \cite{Zoller} and references therein). System
flexibility along with relatively long decoherence times puts it
among top candidates for implementation of quantum information
algorithms \cite{Zoller}. In atomic interferometry \cite{Berman},
ultracold gases in the strongly correlated regime allow to achieve
accuracies below the standard shot noise limit
\cite{Burnett,Rodriquez_interferometry}. A new exciting
application, ``atomtronics'', is suggested by a remarkable analogy
between the physics of ultracold atoms in optical lattices and
that of electrons in crystals \cite{Seaman}. With the current
experimental technique, it seems plausible to produce such basic
atomtronic devises as diodes and bipolar junction transistors,
which serve as building blocks for amplifiers and logic gates.

On the fundamental physics side, strongly interacting lattice
bosons provide insight into the nature of quantum phase
transitions. In particular, these systems are an accurate
experimental realization of the Bose-Hubbard model
\cite{Zoller98},
\begin{equation}
H=-t \sum_{<i,j>} a^{\dagger}_i a^{\:}_j + \frac{1}{2}U \sum_i
\hat{n}_i(\hat{n}_i-1) + \sum_i \epsilon_i \hat{n}_i,
\label{Bose-Hubbard-Hum}
\end{equation}
where $\hat{n}_i=a^{\dagger}_i a_i$ is the number operator on site
$i$; $a^{\dagger}_i$ and $a_i$   respectively create and
annihilate bosons on lattice sites, and $<i,j>$ denotes the sum
over nearest neighbors. The first term describes tunneling between
neighboring potential wells of the optical lattice, the second
term is the effective repulsion within a well, while the last term
is due to an additional smooth space-varying potential, such as,
e.g., a magnetic trap. This system exhibits a transition between
superfluid (SF) and Mott-insulator (MI) groundstates governed by
the competition of atom mobility and interatomic interaction
\cite{Fisher}. The SF-MI phase transition is a topic of intense
current research both theoretically
\cite{Batrouni,Bloch_theor,Isacsson} and experimentally
\cite{Bloch_spatial,Gerbier,Clark}. In typical experiments, a
prepared Bose-Einstein condensate of ultracold atoms is driven to
the strongly correlated regime by means of its adiabatic loading
into a periodic optical potential (optical lattice) induced by the
a.c. Stark effect of interfering laser beams. The mobility of
atoms, namely the hopping $t$, and their interaction $U$ are
controlled by the depth of the optical potential, i.e. by the
laser intensity. In relatively shallow potentials, atoms are
delocalized over the entire lattice giving rise to long-range
coherence and thus \cite{Penrose_Onsager} to superfluid behavior.

If the lattice filling is commensurate, i.e. there is on average
an integer number of atoms per lattice site, increasing the
lattice depth brings the system across the phase transition to the
MI state, which is characterized by zero compressibility and a gap
in the spectrum of elementary excitations. Here, the key
observable is the atom interference pattern obtained upon
releasing the atoms and letting the atom cloud expand for a
transient time of flight \cite{Bloch_SFMI}. Phase correlations
between lattice sites result in pronounced interference peaks
smeared by the finite correlation length in the MI regime. At the
same time phase coherence is fundamentally connected with the
statistics of atom number fluctuations on lattice sites. In
particular, one expects the on-site atom number to behave as a
canonically conjugate variable with respect to the phase field
and, therefore, experience suppressed fluctuations in the MI
regime, analogous to the number squeezed states with
sub-Poissonian number fluctuations \cite{Orzel} widely studied in
quantum optics (see, e.g., \cite{quant_opt}). In practice, number
squeezed states are important for high-precision atomic
interferometry \cite{Berman}, where their use can potentially lead
to sensitivities limited only by the Heisenberg uncertainty
principle \cite{Caves,Burnett,Rodriquez_interferometry}, and for
atom-based quantum computing techniques \cite{Zoller}, where
unwanted number fluctuations necessitate correction procedures in
operation of quantum gates.

Until recently, the number distribution was not measured directly.
The situation changed with the development of the spin-oscillation
technique \cite{Bloch_spin_dynamics}, which is sensitive to the
number of atom pairs and works at arbitrary lattice depths
\cite{Gerbier}, and, most recently, the microwave spectroscopy
using atomic clock shifts \cite{Ketterle_spatial}. In
Ref.~\cite{Gerbier}, Gerbier \textit{et al.} observed a drastic
change of atom number statistics as the system of $^{87}$Rb atoms
was driven through the SF-MI transition. On the theoretical side,
apart from the recent mean-field calculation \cite{Lu-Yu}, a
comprehensive study of atom number fluctuations in the strongly
correlated regime is still missing.

In the present work, we attempt to close this gap by tackling the
problem both analytically and numerically. In
section~\ref{sec:homog}, we focus on the academic case of a
homogeneous square lattice in the thermodynamic limit and in the
limit of zero temperature in one-(1D), two-(2D) and three-(3D)
dimensions. At $U=0$, the ground state of an ideal Bose gas is a
Bose-Einstein condensate with characteristic Poisson distribution
of number fluctuations. In subsection~\ref{subsec:weak_coupling},
we consider the limit of weak interaction $\nu U/t \ll 1$, where
$\nu$ is the filling factor. This parameter region corresponds to
perturbative squeezing of the Poisson distribution, which, as a
function of $\nu U/t$, qualitatively depends on the space
dimension $d$. We consider the case of large filling factors $\nu
\gg 1$ separately in subsection~\ref{subsec:large_nu} because it
is qualitatively different from that of $\nu \sim 1$. Indeed, the
quantum nature of the SF-MI transition implies that the number
fluctuations in the vicinity of the transition must be of order
unity. At the same time, at $U=0$ the variance of the on-site
number distribution is $\sigma^2=\nu \gg 1$. Therefore, there
exists an extensive range of interactions (defined by the
condition $U/t \lesssim \nu$), in which the system remains
superfluid, but its on-site number distribution is drastically
squeezed before the SF-MI can take place. We show that, at $U/t
\ll \nu$, the on-site number statistics are Gaussian and derive
the variance $\sigma^2$ of the distribution, which scales as
$\sigma^2 \propto \sqrt{\nu t / U}$ at $1/\nu \ll U/t \ll \nu $ in
all dimensions. In subsection~\ref{subsec:inter_formula}, we
suggest an expression that interpolates $P(n)$ between the
limiting cases of small interaction and large occupation numbers,
which is found to properly describe $P(n)$ up to $U/t$ of the
order of the critical value $(U/t)_c$. The strong coupling limit,
$\nu t/U \ll 1$, at integer filling is considered in
subsection~\ref{subsec:strong_coupling}. In this limit, the system
is in the MI regime and the on-site number distribution is
governed by rare particle-hole fluctuations.

We study the distribution in the strongly correlated regime
connecting the limiting cases by means of a direct numeric
simulation of the model~(\ref{Bose-Hubbard-Hum}) at $\nu=1$ in
subsection~\ref{subsec:numerics}. The distribution of the on-site
occupation number is a local property, revealing no critical
features at the transition. However, the strongly correlated
region is responsible for the crossover that changes the
statistics qualitatively. As the interaction strength is
increased, we observe a gradual squeezing of the on-site number
distribution and the emergence of the symmetry between particle-
and hole-like fluctuations, characteristic of a MI. In this
section, we also present numerical data for the case of large
filling factors and demonstrate that the analysis of
subsection~\ref{subsec:large_nu} is applicable already at $\nu=5$.

The worm algorithm quantum Monte Carlo (MC) technique
\cite{Worm_Algorithm} allows us to simulate system sizes that are
currently realized in experiments without any approximations,
including the particle number. The results of a direct numeric
simulation of the experimental setup of Ref.~\cite{Gerbier} are
presented in section \ref{sec:exp}. With the lattice parameters
fixed by the experiment, we are left with temperature as the only
free parameter. Due to a smooth confining potential present on top
of the optical lattice, the number distribution is not
homogeneous. We focus on an integral characteristic of the number
distribution, namely, the fraction of atoms found on lattice sites
with occupation $n$, which can be systematically measured
experimentally. This quantity has a pronounced temperature
dependence in both SF and MI regimes.

The problem of thermometry in optical lattices, especially in the
MI regime, is a long standing one. The ability to control the
temperature is of crucial importance for applications that rely on
the peculiar properties of a MI state. At $T=0$ fluctuations are
of quantum nature and can be efficiently controlled externally
through the lattice parameters, whereas temperatures comparable to
the Mott excitation gap destroy the insulating state by activating
particle-hole excitations. At the moment, there are no
experimental techniques to measure the temperature of a strongly
interacting system. Unlike in the weakly interacting regime, where
the temperature can be straightforwardly extracted from the
momentum distribution (e.g., from the interference pattern of
matter waves or from the condensate fraction observed after the
trap is released and the gas expands freely), in the strongly
correlated regime, both temperature and interatomic interaction
are responsible for filling the higher momentum states making
standard absorption imaging techniques inapplicable.

The idea of using occupation number distributions to estimate the
temperature was explored in Ref.~\cite{Prokofev}, where in was
argued that the temperature dependence of the total number of
pairs and their spatial distribution (in traps) provides a
sensitive method of thermometry deeply in the MI phase at energies
smaller than the interatomic interaction, but much larger than the
effective hopping between the sites. [Recently, it has become
possible to directly sample spatially-resolved number
distributions by spin changing collisions \cite{Bloch_spatial},
microwave spectroscopy \cite{Ketterle_spatial}, and the scanning
electron microscope \cite{Gericke_scan_microscp} promising a
complete practical realization of this method.] In this paper, we
perform thermometry of the system in all strongly correlated
regimes by comparing experimental data and numerical results for
the statistics of occupation numbers. More specifically, we
compare numerically calculated fraction of pairs ($n=2$) with that
measured in Ref.~\cite{Gerbier} across the SF-MI transition
estimating the range of experimental temperatures. The accuracy of
this method is mainly limited by the error bars of the
experimental data and by the range of applicability of the
Bose-Hubbard model.

%%%%%%%%%%%%%%%%%%%%%%%%%%%%%%%%%%%
\section{ Homogeneous Lattice} \label{sec:homog}
%%%%%%%%%%%%%%%%%%%%%%%%%%%%%%%%%%%

In this section, we assume that there is no space-varying
potential on top of the optical lattice, and set $\epsilon_i
\equiv \epsilon_0$. Let $N_s$ be the number of lattice sites and
$N$ be the total number of particles. The goal of this section is
to obtain the ground state probability $P_n$ to detect $n$
particles on a given lattice site in the limit of $N_s,N
\rightarrow \infty$, at a fixed filling factor $\nu=N/N_s$.
Mathematically, $P_n$ can be defined as
\begin{equation}
P_n (U/t)= \sum_{\{n_{i \ne 1}\} }\big|\; \langle n_1=n, \{n_{i
\ne 1}\} | \:|\Psi_{U/t} \rangle \big| ^2 \; , \label{Pn}
\end{equation}
where $| \Psi_{U/t} \rangle$ is the many-body ground state
wavefunction, and  $| \{n_{i}\} \rangle$ are Fock states.

%%%%%%%%%%%%%%%%%%%%%%%%%%%%%%%%%%%%%%%%%%%%%%%%%%%%%
\subsection{Weak coupling limit} \label{subsec:weak_coupling}
%%%%%%%%%%%%%%%%%%%%%%%%%%%%%%%%%%%%%%%%%%%%%%%%%%%%%

At sufficiently small $U/t$, the relevant representation of
Eq.~(\ref{Bose-Hubbard-Hum}) is obtained by the diagonalization of
the kinetic energy term with the canonical transformation
$a_i=N_s^{-1/2}\sum_\mathbf{k} b_\mathbf{k} \exp(i 2\pi\:
\mathbf{k} \, \mathbf{r}_i/L)$ (periodic boundary conditions are
assumed), where $L=N_s^{1/d}$ is the linear system size,
$\mathbf{r}_i$ and $\mathbf{k}$ are respectively the position of
the site $i$ and a quasi-momentum in $d$ dimensions with
integer-valued components, $-(L-1)/2 \leq r_{i\: \mu},\: k_{\mu}
\leq (L-1)/2$, $\mu=1,\, ... \, ,d$. The result is
\begin{equation}
H=\sum_{k} \varepsilon_{\mathbf{k}} b^{\dagger}_{\mathbf{k}}
b^{\;}_{\mathbf{k}} + \frac{U}{2N_s} \sum_{\mathbf{k}_1 +
\mathbf{k}_2= \mathbf{k}_3 + \mathbf{k}_4}
b^{\dagger}_{\mathbf{k}_1}b^{\dagger}_{\mathbf{k}_2}
b^{\:}_{\mathbf{k}_3}b^{\:}_{\mathbf{k}_4}
\label{Bose-Hubbard-Hum_momentum_space}
\end{equation}
with $\varepsilon_{\mathbf{k}}\, =\,
2t\sum_{\mu=1}^{d}\,[1-\cos(2\pi k_{\mu}/L)]$. At $U=0$ the ground
state of the Hamiltonian (\ref{Bose-Hubbard-Hum_momentum_space})
is a pure Bose-Einstein condensate, which can be expressed as a
coherent state, $ |\Psi_{(U/t)=0}\rangle = \exp(\sqrt{N}
b_0^{\dagger} - N/2) | 0 \rangle$. Then, transforming the
wavefunction back to the on-site representation in terms of $\{
a_i \}$ yields the Poisson distribution for the probability to
find $n$ particles on a given site:
\begin{equation}
P^{(0)}_n=e^{-\nu} \, \frac{{\nu}^{n}}{n!} \, .
\label{Pn_Ideal_gas}
\end{equation}
At a finite, but small, $U$ we employ the standard Bogoliubov
method \cite{LLStatMech2} of separating the system into the
classical-field condensate part and non-condensate particles
interacting with it, omitting the terms of the third and forth
order with respect to the non-condensate operators. In this
approximation, the Hamiltonian
(\ref{Bose-Hubbard-Hum_momentum_space}) is reduced to a bilinear
in $b_{\mathbf{k}}$ and $b_{\mathbf{k}}^{\dagger}$ form and
diagonalized by the canonical transformation
$c_{\mathbf{k}}=u_\mathbf{k} b_{\mathbf{k}} + v_\mathbf{k}
b_{-\mathbf{k}}^{\dagger}$, where
\begin{gather}
u_\mathbf{k} = \left[(\varepsilon_\mathbf{k} + \nu U)/2\omega_\mathbf{k} + 1/2 \right]^{1/2}, \notag \\
v_\mathbf{k} = \left[(\varepsilon_\mathbf{k} + \nu U)/2\omega_\mathbf{k} - 1/2 \right]^{1/2}, \notag \\
\omega_\mathbf{k}=\left[\varepsilon_\mathbf{k}^2 + 2 \nu
\varepsilon_\mathbf{k} U\right]^{1/2}. \label{UkVk}
\end{gather}
The ground-state wavefunction is then obtained from the equation
$c_{\mathbf{k}}\:|\Psi_{U/t} \rangle = 0$ for all ${\mathbf{k}}
\neq 0$ and has the form
\begin{equation}
|\Psi_{U/t}\rangle = C \exp\Biggl[\sqrt{N_0} b_0^{\dagger} -
\frac{N_0}{2} - \frac{1}{2}
\sum_{\mathbf{k}\neq0}\frac{v_\mathbf{k}}{u_\mathbf{k}}b^{\dagger}_{\mathbf{k}}b^{\dagger}_{-\mathbf{k}}\Biggr]
| 0 \rangle, \label{Psi_Bogolubov}
\end{equation}
where $C$ is the normalization factor and $N_0=N -
\sum_{\mathbf{k}\neq0} \langle
b^{\dagger}_{\mathbf{k}}b_{\mathbf{k}}\rangle$ is the number of
condensate particles.

Now we can express $|\Psi_{U/t}\rangle$ in terms of the on-site
operators,
\begin{gather}
|\Psi_{U/t}\rangle = C \exp\Biggl[ \sum_i \left(\sqrt{\nu_0}
a_i^{\dagger} - \frac{\nu_0}{2} \right) - \sum_{i,j} S_{ij}
a^{\dagger}_i a^{\dagger}_j \Biggr] | 0 \rangle \;,
\label{Psi_onsite}
\end{gather}
with $\nu_0=N_0/N_s$ and
\begin{equation}
S_{i j} =   \frac{1}{2N_s}
\sum_{\mathbf{k}\neq0}\frac{v_\mathbf{k}}{u_\mathbf{k}} e^{i 2 \pi
\mathbf{k} (\mathbf{r}_i - \mathbf{r}_j)/L}. \label{Sij}
\end{equation}
In this form, the wavefunction can be used to obtain the on-site
number distribution in the whole range of $U/t \ll \nu$ by a
straightforward application of Eq.~(\ref{Pn}).

We derive a closed-form expression for $P(n)$ in the limiting case
of
\begin{equation}
\alpha \, = \, \nu \, U/t \, \ll 1, \label{alpha}
\end{equation}
which corresponds to the range of sufficiently weak squeezing of
$P(n)$ allowing us to consider only the first correction to the
Poisson distribution in the leading power of $\alpha$.
Mathematically, Eq.~(\ref{alpha}) implies that $S_{ij} \ll 1$. If,
in addition, $S_{ij}$ is short-range, i.e. it decays at distances
$|\mathbf{r}_i - \mathbf{r}_j| \sim 1$, which implies that the
leading correction is insensitive not only to the system size, but
also to the value of the healing length $ \propto \alpha^{-1/2}$,
then we can expand the exponential in Eq.~(\ref{Psi_onsite}) in
powers of $S_{ij}$. Rather straightforward but lengthy algebra
yields the distribution in the form
\begin{equation}
P_n=P^{(0)}_n- \frac{\nu \,\lambda(\alpha)}{2} \,
[P^{(0)}_n-2P^{(0)}_{n-1}+P^{(0)}_{n-2}] \:,
\label{Pn_weak_coupling}
\end{equation}
where $\lambda$ is an interaction-dependent squeezing parameter
and $P^{(0)}_n$ is given by Eq.~(\ref{Pn_Ideal_gas}) assuming
$P^{(0)}_n=0$ for $n < 0$. Identically,
\begin{equation}
P_n = P^{(0)}_n \left(1 + \frac{\lambda(\alpha)}{2} \, \left[
\frac{n - (n-\nu)^2}{\nu}\right] \right) . \label{Pn_relative}
\end{equation}

Let us postpone writing an explicit expression for $\lambda$ and
discuss the underlying assumptions leading to
Eqs.~(\ref{Pn_weak_coupling}),(\ref{Pn_relative}). It turns out
that $S_{ij}$ is local only in 3D, where the main contribution to
the integral in Eq.~(\ref{Sij}) comes from large momenta close to
the edge of the Brillouin zone, $|\mathbf{k}| \sim L/2$. In 2D,
the integral has a logarithmic divergency at low momenta in the
limit $\alpha \to 0$, while, in 1D, the dominant contribution to
$S_{ij}$ comes from small momenta, meaning that in both 1D and 2D
cases we can not rely on the perturbative expansion of the wave
function. Nevertheless,
Eqs.~(\ref{Pn_weak_coupling}),(\ref{Pn_relative}) are still valid
in 1D and 2D, since the functional form of $P(n)$ must be the same
in all dimensions. To prove this statement, we note that $P(n)$ is
unambiguously determined by its characteristic function
$\chi(t)=\langle \exp(i \, t \, a_i^{\dagger} a_i) \rangle$, which
can be used to generate all moments of $P(n)$. The function
$\chi(t)$ can be calculated explicitly as a series expansion in
powers of $(i t a_i^{\dagger}a_i)$. Since averages of bosonic
quasiparticle operators in the Bogoliubov theory obey Wick's
theorem, each term in the series is a function of $\zeta \equiv
\langle a_i^{\dagger} a_i \rangle$ and $\zeta' \equiv \langle a_i
a_i \rangle$. Therefore, $\chi(t) = \chi(t, \zeta, \zeta')$,
meaning that all physical parameters, including the space
dimension, enter $P(n)$ only through $\zeta$ and $\zeta'$. In
Eqs.~(\ref{Pn_weak_coupling}),(\ref{Pn_relative}), this
combination determines $\lambda(\alpha)$.

To determine $\lambda(\alpha)$ in all dimensions were note that it
is directly related to the dispersion of $P(n)$ in
Eq.~(\ref{Pn_relative})
\begin{equation}
\sigma^2 \equiv \langle (n-\nu)^2 \rangle = \nu \, (1-\lambda).
\label{sigma-lambda}
\end{equation}
On the other hand, $\sigma^2 = \langle n_i^2 \rangle - \nu^2 =
\langle a_i^{\dagger}a^{\;}_i a_i^{\dagger}a^{\;}_i \rangle -
\nu^2$ can be calculated in a standard way by replacing $a_i$ with
their expressions in terms of the Bogoliubov modes and the
classical-field condensate part, i.e.
$a_i=\sqrt{\nu_0}+N_s^{-1/2}\sum_{\mathbf{k} \neq 0} [u_k
c_\mathbf{k}-v_k c^{\dagger}_{-\mathbf{k}}] \exp(i 2\pi\:
\mathbf{k} \, \mathbf{r}_i/L)$. Then, an application of the Wick's
theorem along with $\langle c_{\mathbf{k}}
c^{\dagger}_{\mathbf{k}'} \rangle =
\delta_{\mathbf{k},\mathbf{k}'}$ and $\langle c_{\mathbf{k}}
c_{\mathbf{k}'} \rangle = 0$ gives
\begin{equation}
\sigma^2=\frac{\nu}{N_s} \sum_{\mathbf{k} \neq 0}
\frac{\varepsilon_{\mathbf{k}}}{\omega_{\mathbf{k}}} \; .
\label{sigma-general-zeroT}
\end{equation}

Strictly speaking, Eq.~(\ref{sigma-general-zeroT}) is valid as
long as the number of non-condensed particles is small (the 1D
case is special in this regard and is further discussed below),
$(N-N_0)/N \ll 1$, which implies $U/t \ll \nu^{(2-d)/d}$ at small
$\nu$ and $U/t \ll \nu$ at large $\nu$. In the latter case, the
distribution change can be non-perturbative since $\lambda \sim 1$
is typical in this parameter regime, which will be discussed in
more detail in the next subsection.

Thus, we arrive at the following result
\begin{equation}
\lambda=\frac{1}{N_s} \sum_{\mathbf{k} \neq 0} \big[ 1-
\frac{\varepsilon_{\mathbf{k}}}{\omega_{\mathbf{k}}} \big] \; .
\label{lambda-general}
\end{equation}
The asymptotic behavior of $\lambda(\alpha \to 0)$ qualitatively
depends on the dimensionality. In 1D, the main contribution to the
integral (\ref{lambda-general}) comes from low momenta resulting
in
\begin{equation}
\lambda_{(d=1)} (\alpha) \, \to\,  \frac{\sqrt{2}}{\pi}\,
\sqrt{\alpha}. \label{lambda_1D}
\end{equation}
In 3D, there is no low-momentum singularity at $\alpha \to 0$, and
the asymptotic expression is linear in $\alpha$:
\begin{equation}
\lambda_{(d=3)} (\alpha) \, \to \,  \frac{B}{2 \pi^3}\; \alpha,
\label{lambda_3D}
\end{equation}
\begin{equation}
B\,=\, \int_0^{\pi}\!\!\! \int_0^{\pi}\!\!\!\int_0^{\pi}
\frac{dz_1\, dz_2\, dz_3}{\sum_{\mu=1}^3 (1-\cos z_{\mu})}\,
\approx \, 15.673. \label{B}
\end{equation}
In 2D, the low-momentum singularity is logarithmic
\begin{equation}
\lambda_{(d=2)} (\alpha) \; \to \; {\ln (C/\alpha)\over\ 4 \pi}\;
\alpha, \;\;\;\;\; C\, \approx\, 23.54  \label{lambda_2D}
\end{equation}

A comment is in order here concerning the derivation procedure for
the 1D case. Formally, the condensate fraction is zero in the
macroscopic limit even at $T=0$, while the derivation is based on
the assumption that almost all the particles are Bose condensed.
Nevertheless, our final results for the probabilities are valid
even in 1D, and the generic reasoning---based on the notion of
quasicondensate---leading to this conclusion is as follows
\cite{KSS}. The quasicondensate correlation properties
characteristic of a weekly interacting 1D gas at $T=0$ imply two
different correlation radii, $r_c$ and $R_c$, $R_c \gg r_c$. Here
$r_c$ defines the length scale upon which the system can be
considered as macroscopic, while $R_c$ is the length at which
(quantum) fluctuations of phase are of order unity. If the system
size $L$ is such that
\begin{equation}
r_c \ll L \ll R_c \; , \label{L}
\end{equation}
then the system is macroscopic with respect to all {\it local}
properties, while still featuring a genuine condensate. (In a 1D
weakly interacting system at $T=0$ the density of this condensate
is close to the total density of particles.) Hence, for all local
properties, including the ones discussed in the present paper, one
can assume, without loss of generality, that the system size is
finite and satisfies the condition (\ref{L}). It should be
emphasized that this assumption is {\it implicit} and is used
exclusively to justify the derivation procedure. Otherwise, it
does not lead to any explicit dependence of final answers on $L$.
Indeed, the first inequality in Eq.~(\ref{L}) guarantees that all
discrete sums over momenta can be replaced with integrals, and,
since the integrals are convergent, the answer is independent of
$L$.

%%%%%%%%%%%%%%%%%%%%%%%%%%%%%%%%%%%%%%%%%%%%%%%%%%%%%
\subsection{Large occupation number limit} \label{subsec:large_nu}
%%%%%%%%%%%%%%%%%%%%%%%%%%%%%%%%%%%%%%%%%%%%%%%%%%%%%

At $\nu \gg 1$ there exists a wide ($U/t \ll \nu$) superfluid
region, in which large number fluctuations given by
Eq.~(\ref{Pn_Ideal_gas}) are gradually suppressed by the
interaction until they become of order unity at the SF-MI
transition. This physically appealing regime is not captured by
Eqs.~(\ref{Pn_weak_coupling}),(\ref{Pn_relative}), since they are
applicable only for $ U/t \ll 1/\nu$.

At $\nu \gg 1$ and $U/\nu t \ll 1$, the number distribution  is
easily obtained due to the fact that the typical values of the
occupation number fluctuations, $\eta_i = a_i^{\dagger}a_i - \nu$,
are large $1 \ll |\eta_i| \ll \nu$. Thus, the transformation
$a_i=\sqrt{\nu + \eta_i}\exp(i \Phi_i)$, where $\eta_i$ and
$\Phi_i$ are canonically conjugate Hermitian operators, along with
$\eta_i/\nu \ll 1$ reduces the Bose-Hubbard
Hamiltonian~(\ref{Bose-Hubbard-Hum}) to
\begin{multline}
H=-t \nu \sum_{<i,j>} \left[ 1 +
\frac{1}{4\nu^2}(\eta_i\eta_j-\eta_i^2)\right]\cos(\Phi_i-\Phi_j)
\\ + \frac{U}{2}\sum_i \eta_i^2, \label{Extended-Rotor-cos-Hum}
\end{multline}
where $0 \leq \Phi_i \leq 2 \pi$. In fact, in this form,
Eq.~(\ref{Extended-Rotor-cos-Hum}) is applicable at {\it any}
$\eta_i \ll \nu$ including the description of the SF-MI transition
and the MI phase. At sufficiently large interaction, namely $U/t
\gg 1/\nu$, the $\propto \eta_i^2/\nu^2$ term in
Eq.~(\ref{Extended-Rotor-cos-Hum}) can be neglected, and the
Hamiltonian coincides with the well-studied quantum rotor model
(QRM).

In this work, we are interested in the $U/ t \ll \nu$ limit of
model Eq.~(\ref{Extended-Rotor-cos-Hum}), i.e. when phase
fluctuations between the nearest-neighbor sites are small $|\Phi_i
-\Phi_j| \ll 1$ and the number fluctuations are large. This allows
us to consider $\eta_i$ as a continuous variable and formally
redefine the domain of $\eta_i,\Phi_i$ as $-\infty < \eta_i,
\Phi_i < \infty$. The result is a harmonic approximation of the
Hamiltonian~(\ref{Extended-Rotor-cos-Hum}),
\begin{multline}
H=t \nu \sum_{<i,j>} \left[ \frac{1}{2} (\Phi_i-\Phi_j)^2 +
\frac{1}{4\nu^2}(\eta_i^2-\eta_i\eta_j)\right] \\ +
\frac{U}{2}\sum_i \eta_i^2.~~~~~~~~~ \label{Extended-Rotor-Hum}
\end{multline}

From its functional (quadratic) form we immediately conclude that
the distribution of $\eta_i$, which is Gaussian at $U=0$ ($\nu \to
\infty$ limit of Eq.~(\ref{Pn_Ideal_gas})), remains Gaussian in a
wide range of coupling strength---all the way to the vicinity of
the SF-MI transition where $\sigma^2 \sim 1$ and the
model~(\ref{Extended-Rotor-Hum}) breaks down. The proof is as
follows. Since the Hamiltonian~(\ref{Extended-Rotor-Hum}) is
bilinear in $\{\eta_i, \Phi_i\}$, all averages are subject to the
Wick's theorem. Therefore, the characteristic function of the
distribution $W(\eta_i)$, which is given by the integral
$\int_{-\infty}^{\infty} \exp(ik\eta_i) W(\eta_i) \: d \eta_i$, is
a Gaussian $\exp(-k^2 \sigma^2/2)$.

The only parameter of the distribution, $\sigma^2$, is given by
\begin{equation}
\sigma^2=\frac{\nu}{N_s} \sum_{\mathbf{k}}
\frac{\varepsilon_{\mathbf{k}}}{\omega_{\mathbf{k}}} \; \left[\,
1+ 2\langle c_{\mathbf{k}}^{\dagger}c_{\mathbf{k}} \rangle \,
\right], \label{sigma-general}
\end{equation}
where $c_{\mathbf{k}}$ and $c_{\mathbf{k}}^{\dagger}$ are the
creation and annihilation operators of the normal modes of the
Hamiltonian~(\ref{Extended-Rotor-Hum}) with frequencies
$\omega_\mathbf{k}$ given by Eq.~(\ref{UkVk}). The Hamiltonian is
diagonalized by the transformation
\begin{gather}
\eta_i = \sum_\mathbf{k}
\sqrt{\frac{\omega_\mathbf{k}}{2U+\varepsilon_{\mathbf{k}}/\nu}}
\; \left[ \; \chi_{\mathbf{k} \, i} \, c_\mathbf{k} +
\chi^*_{\mathbf{k} \, i}\, c^{\dagger}_\mathbf{k} \; \right],
\notag
\\ \Phi_i = - i \sum_\mathbf{k} \sqrt{\frac{\omega_\mathbf{k}}{2\nu\varepsilon_{\mathbf{k}}}} \; \left[ \; \chi_{\mathbf{k} \, i} \, c_\mathbf{k}
- \chi^*_{\mathbf{k} \, i}\, c^{\dagger}_\mathbf{k} \; \right],
\label{phonon_fields} \\
\chi_{\mathbf{k} \, i}=\frac{1}{\sqrt{N_s}}\exp(i 2\pi\:
\mathbf{k} \, \mathbf{r}_i/L) \notag
\end{gather}

Here, we restrict ourselves to ground state properties only and
thus set $\langle c_{\mathbf{k}}^{\dagger}c_{\mathbf{k}} \rangle
\equiv 0$, which leads exactly to Eq.~(\ref{sigma-general-zeroT}).
This is hardly surprising, since the
model~(\ref{Extended-Rotor-Hum}) is equivalent to the Bogoliubov
approximation of the
Hamiltonian~(\ref{Bose-Hubbard-Hum_momentum_space}) in the limit
of large $\nu$, and thus, we could formally demonstrate that
$P(n)$ is Gaussian at $\nu \gg 1$ already in the framework of
section~\ref{subsec:weak_coupling}. However, the hydrodynamic
approach chosen in this section seems more natural and physically
transparent when dealing with a dense system.

Let us examine properties of the distribution variance in more
detail. An explicit expression for $\sigma^2$ reads
\begin{equation}
\sigma^2=\frac{\nu}{\pi^d} \int_0^\pi \cdots \int_0^\pi
\sqrt{\frac{\sum_\mu (1-\cos x_\mu)}{\nu U/t + \sum_\mu (1-\cos
x_\mu)}} \: d x_1 \cdots d x_d, \label{sigma-zeroT}
\end{equation}
In the limit of $\alpha=\nu U/t \rightarrow 0$, this expression
reduces to Eqs.~(\ref{lambda_1D})-(\ref{lambda_2D}). In the
opposite limit, we find
\begin{equation}
\sigma^2= \frac{I_d}{\pi^d} \sqrt{\frac{\nu t}{U}},~~~~~~~ U/t \gg
1/\nu,\label{sigma-largeU}
\end{equation}
where $I_1 = 2 \sqrt{2}$, $I_2 \approx 13.373$, and $I_3 \approx
52.348$. The latter formula allows to estimate the range of $U/t$
at which Eq.~(\ref{sigma-zeroT}) is applicable. The condition
$\sigma^2 \gg 1$ gives $U/t \ll \nu$, consistent with the
applicability range of Eq.~(\ref{sigma-general-zeroT}).

%%%%%%%%%%%%%%%%%%%%%%%%%%%%%%%%%%%%%%%%%%%%%%%%%%%%%
\subsection{Interpolation formula in the SF regime} \label{subsec:inter_formula}
%%%%%%%%%%%%%%%%%%%%%%%%%%%%%%%%%%%%%%%%%%%%%%%%%%%%%
In sections~\ref{subsec:weak_coupling} and \ref{subsec:large_nu},
we have derived asymptotically exact solutions describing week
squeezing ($\lambda \ll 1$) and strong squeezing ($0 \leq \lambda
\lesssim 1$) at $\nu \gg 1$ respectively. The two limits overlap,
which suggests that one can write a single interpolation formula
to capture both limits correctly. This formula is also expected to
predict $P(n)$ correctly in a broader region of $U/t$ even at $\nu
\sim 1$. Formally, we have to find a function $\tilde{P}(n, \nu,
\lambda)$ such that (i) it coincides with Eq.(\ref{Pn_relative})
in the limit of $\lambda \ll 1$, (ii) it becomes Gaussian as a
function of $n$ at $\nu \gg 1$ and $0 \leq \lambda <1$ with the
average $\nu$ and variance $\tilde{\sigma}^2=\nu(1-\lambda)$, and
(iii) it is analytic with respect to $\nu$ and $\lambda$. The
solution is not unique, and we simply suggest one which satisfies
the above mentioned criteria
\begin{multline}
\tilde{P}\,(n,\nu,\lambda) = c \, P^{(0)}(n) \\
\exp\left[\frac{\lambda (n-\nu) + (n- \nu^2)}{2\nu} -
\frac{(n-\nu)^2}{2 \nu (1-\lambda)}\right],
\label{Pn-tilde-general}
\end{multline}
where $c$ is the normalization factor and $P^{(0)}(n)$ is the
Poisson distribution. By comparing Eq.~(\ref{Pn-tilde-general})
with numerical simulations at $\nu=1$ (described below) we find
that this formula accurately describes the actual form of $P(n)$
in a much broader range of $U/t$ than the
solution~(\ref{Pn_relative}).

%%%%%%%%%%%%%%%%%%%%%%%%%%%%%%%%%%%%%%%%%%%%%%%%%%%%%
\subsection{Strong coupling limit} \label{subsec:strong_coupling}
%%%%%%%%%%%%%%%%%%%%%%%%%%%%%%%%%%%%%%%%%%%%%%%%%%%%%

Let us now turn to the strong coupling limit, $\nu t / U \ll 1$,
at integer $\nu$, when atoms are well in the Mott insulator
regime, and the zeroth order approximation to the wavefunction is
a direct product of local Fock states, $|\Psi_{(t/U)=0}\rangle =
|\{ n_i=\nu \} \rangle$. The effect of finite hopping $t$ can be
taken into account as a perturbation of the on-site interaction
term in the Hamiltonian~(\ref{Bose-Hubbard-Hum}). To the first
approximation, this results in an admixture of the particle-hole
pairs to $|\Psi_{(t/U)=0}\rangle$,
\begin{equation}
|\Psi_{t/U}\rangle \propto \Bigl(
1+\frac{t}{U}\sum_{<i,j>}a_i^{\dagger}a_j
\Bigr)|\Psi_{(t/U)=0}\rangle, \;\;\; \nu t/ U \ll 1.
\label{Psi_Mott}
\end{equation}
With this wavefunction  Eq.~(\ref{Pn}) leads to the following
distribution
\begin{gather}
P_{\nu}=1-2P_{\nu-1}, \notag \\
P_{\nu-1}=P_{\nu+1}=2d \, \nu(\nu+1) \; t^2/U^2, \notag \\
P_n= 0, \; \mathrm{if} \;|n-\nu| > 1 . \label{Pn_Mott}
\end{gather}

%%%%%%%%%%%%%%%%%%%%%%%%%%%%%%%%%%%%%%%%%%%%%%%%%%%%%
\subsection{Numerics} \label{subsec:numerics}
%%%%%%%%%%%%%%%%%%%%%%%%%%%%%%%%%%%%%%%%%%%%%%%%%%%%%

\begin{figure*}[htb]
\includegraphics[width=0.95\textwidth, keepaspectratio=true]{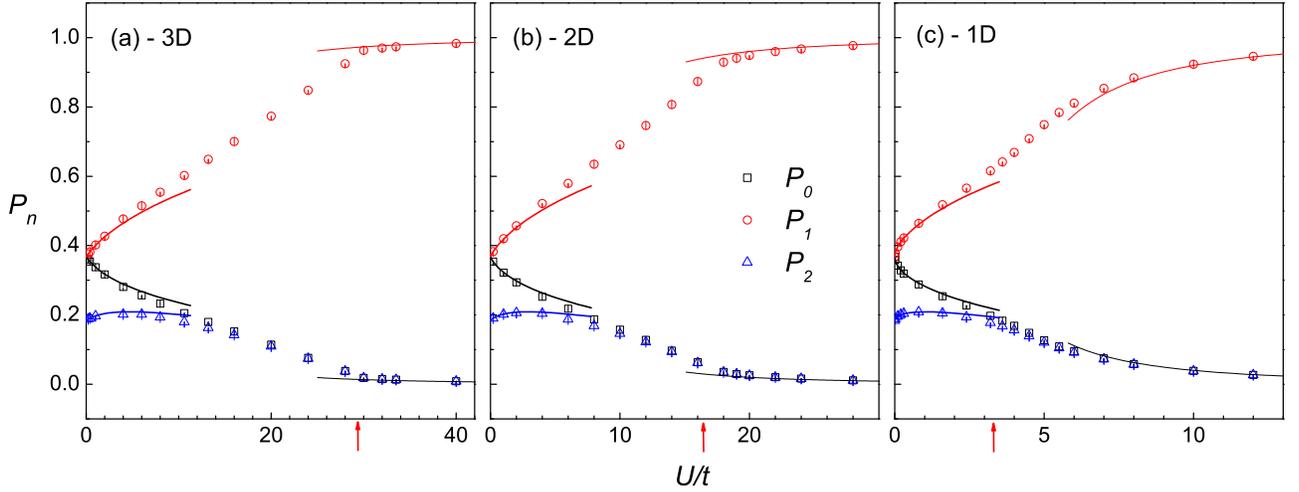}
\caption{(Color online.) The probability $P_n$ to detect $n$
particles on a single site of a homogenous square lattice as a
function of $U/t$ at zero temperature and unitary filling. The
results of MC simulation are shown for $n=0$ (squares), $n=1$
(circles) and $n=2$ (triangles). The uncertainties in $P_n$ are
smaller than the symbol size. The predictions of
Eq.~(\ref{Pn-tilde-general}) in the SF regime and that of
Eq.~(\ref{Pn_Mott}) in the strong coupling limit are plotted by
thick and thin solid lines respectively. The critical points of
the SF-MI transition, marked by the arrows on the graph, are
$(U/t)_c = 29.34(2)$ in three dimensions \cite{Barbara} (a),
$(U/t)_c=16.4(8)$ in two dimensions \cite{Krauth} (b), and
$(U/t)_c=3.30(2)$ in one dimension \cite{Svistunov_1DSFMI} (c).}
\label{fig:homog_case}
\end{figure*}

Clearly, as $U$ is changed from $0$ to $\infty$ the number
distribution must change qualitatively. Having an essentially long
tail at $n > \nu$ in the weakly interacting limit, $P_n$ becomes
sharply peaked at large $U$ with equal probabilities for $\nu-1$
and $\nu+1$ particles on a site. To obtain $P_n(U/t)$ in the
crossover regime, $t \sim U$, we perform MC simulations using the
continuous-time Worm Algorithm scheme \cite{Worm_Algorithm}. We
set $\nu=1$ and take the limit $N_s \rightarrow \infty$,
$\beta=1/T \rightarrow \infty$, where $T$ is the temperature. For
the linear system size $L=N_s^{1/d}=24$ and $\beta \gg L/2\pi c$,
where $c \sim 6t$ is the typical value of sound velocity in the
superfluid phase (higher temperatures can be used in the MI phase
with gaps $\sim U$), the shape of the distribution is already well
saturated, within a fraction of one percent in 3D and 2D, and
within a few percent in 1D, consistent with the fact that $P_n$ is
an essentially local characteristic.

The simulation results for 3D, 2D and 1D are shown in
Fig.~\ref{fig:homog_case}, where $P_0$, $P_1$ and $P_2$ are
plotted as functions of $U/t$ along with the predictions of
Eq.~(\ref{Pn-tilde-general}) and Eq.~(\ref{Pn_Mott}). The main
observation is that, although $P_n(U/t)$ fundamentally does not
reveal any critical behavior, in 3D and 2D the SF-MI transition is
marked by a substantial change in $P_n$ curves---they rapidly
plateau in the Mott regime for $U/t \gtrsim (U/t)_c$. As expected,
in 1D the curves are much more smooth and the saturation at high
$U$ is not pronounced. Remarkably, Eq.~(\ref{Pn-tilde-general}),
deviates notably (by a few percent) from the numerical data only
at $U/t$ as high as $\sim 2$ in 1D, $\sim 4$ in 2D, and $\sim 8$
in 3D.

\begin{figure}[tbh]
\includegraphics[width=0.95\columnwidth,keepaspectratio=true]{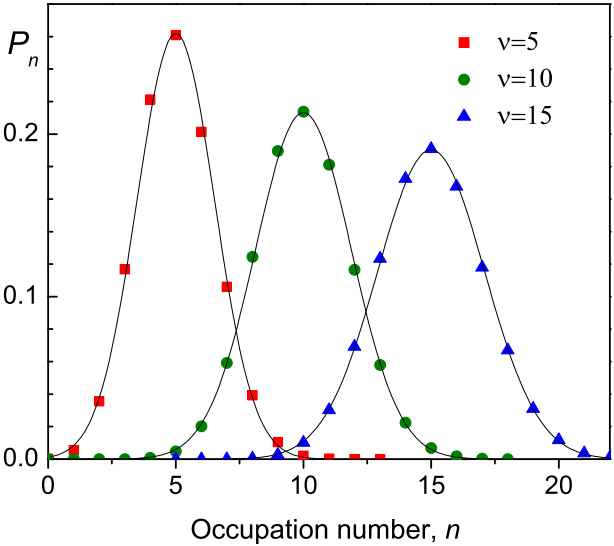}
\caption{(Color online.) Probability $P_n$ of the on-site
occupation number $n=\eta_i+\nu$ at $U/t=2$ from the Monte-Carlo
simulation (the error bars are smaller then the symbol size) with
the filling factor $5$ (squares), $10$ (circles), and $15$
(triangles). The solid lines are the Gaussian curves with the
variances calculated from Eq.~(\ref{sigma-zeroT}) with the
parameters of simulations.}
\label{fig:variance}
\end{figure}

\begin{figure}[tbh]
\includegraphics[width=0.95\columnwidth,keepaspectratio=true]{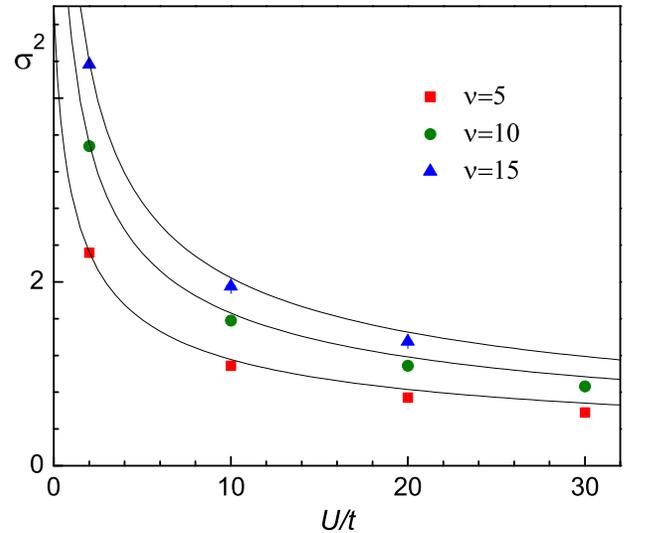}
\caption{(Color online.) The variance $\sigma^2$ as a function of
interaction strength $U/t$ from the Monte-Carlo simulation (the
error bars are smaller then the symbol size) with the filling
factor $5$ (squares), $10$ (circles), and $15$ (triangles). The
solid lines represent the prediction of Eq.~(\ref{sigma-zeroT})
with the parameters of simulations.}
\label{fig:variance_of_U}
\end{figure}

We compare the theoretical predictions of
section~\ref{subsec:large_nu} for the asymptotics in the $\nu \gg
1$ limit with the results of Monte-Carlo simulations. From
Fig.~\ref{fig:variance}, where the simulated probability
distribution at $U/t=2$ is plotted along with Gaussian curves with
variances calculated by Eq.~(\ref{sigma-zeroT}) at corresponding
$\nu$, we conclude that the shape of the distribution is perfectly
well (within the error bars) described by the Gaussian
distribution already at $\nu=5$. In Fig.~\ref{fig:variance_of_U},
the numerically simulated $\sigma^2$ and the prediction of
Eq.~(\ref{sigma-zeroT}) are plotted together as a function of
$U/t$.

%%%%%%%%%%%%%%%%%%%%%%%%%%%%%%%%%%%
\section{Comparison with experiment: effects of finite temperature}
\label{sec:exp}
%%%%%%%%%%%%%%%%%%%%%%%%%%%%%%%%%%%
In actual experiments, atoms are confined in the lattice by  a
trapping potential, typically of a parabolic form. This results in
an inhomogeneous density profile in the SF regime and in the
formation of Mott-plateaus \cite{Gerbier, Bloch_spatial,
Ketterle_spatial}---spherical shells of integer filling---in deep
lattices. Correspondingly, the distribution of the number
fluctuations is also inhomogeneous, i.e. $P_n=P_n(i)$, where $i$
is the site index. The development of experimental techniques
allowing detection of atoms with a single-site spatial resolution
\cite{Gericke_scan_microscp} should open an exciting possibility
to directly measure atom correlation functions and $P_n(i)$ in
particular. Current experiments \cite{Gerbier, Bloch_spatial,
Ketterle_spatial} typically deal with integral characteristics of
the number distribution, such as the fraction of the total number
of atoms found on lattice sites with occupation $n$,
\begin{equation}
f_n = \sum_i n P_n(i). \label{fn}
\end{equation}

The fraction of pairs $f_2$ in both SF and MI regimes can be
accurately probed by the spin-changing collisions
\cite{Bloch_spin_dynamics}. The measurement is set up in the
following way \cite{Gerbier}. After the system is allowed to
equilibrate at a fixed value of the lattice potential depth $V_0$,
the configuration of atoms is frozen by a rapid increase of $V_0$.
Then, coherent spin dynamics in the two-particle channel can be
induced with a near-unitary efficiency, and the spin oscillation
amplitude is measured to yield the fraction of pairs. With this
technique, Gerbier \textit{et al.} \cite{Gerbier} observed $f_2$
as the system was driven from a SF regime across the transition
deep into MI regime, corresponding to the values of the initial
lattice potential $V_0$ ranging from $4 E_r$ to $40 E_r$, where
$E_r=h^2/2m \lambda^2$ is the single photon recoil energy, $m$ is
the mass of a $^{87}$Rb atom, and $\lambda$ is the lattice laser
wavelength.

\begin{figure}[tbh]
\includegraphics[width=1\columnwidth,keepaspectratio=true]{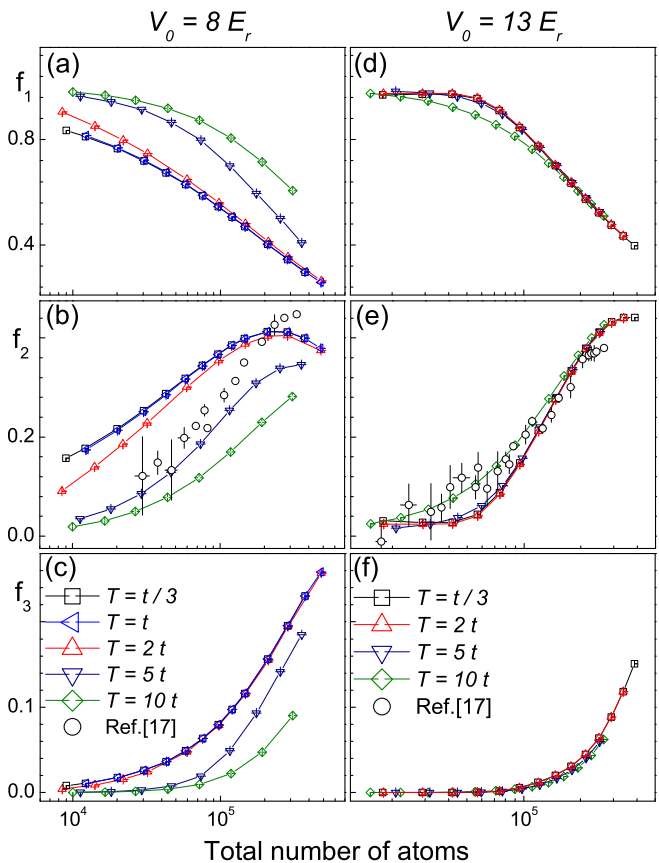}

\caption{(Color online.) Fractions of atoms $f_1$, $f_2$, $f_3$
occupying lattice sites with one, two and three particles on them
respectively versus the total number of particles in the trap at
the lattice depths $V_0=8E_r$, (a), (b), (c), and $V_0=13E_r$,
(d), (e), (f). The numerical results at different temperatures are
plotted with the experimental data for $f_2$ of
Ref.~\cite{Gerbier}.}
\label{fig:exp}
\end{figure}

In our simulations, the system of $^{87}$Rb atoms of
Ref.~\cite{Gerbier} is described by the Bose-Hubbard
model~(\ref{Bose-Hubbard-Hum}). The external potential is
harmonic, $\epsilon_i = m \omega_0^2 \, \mathbf{r}_i^2/2$, and all
the parameters of the Hamiltonian~(\ref{Bose-Hubbard-Hum}) are
fixed by the experimental setup. We study the number fluctuations
at two values of the lattice depth, $V_0 = 8 E_r$ and $V_0 = 13
E_r$ (corresponding to $U/t \approx 7.4$ and $U/t \approx 35.6$
respectively), which serve as examples of typical SF and MI phases
in the strongly correlated regime. The trapping frequency
$\omega_0$ is equal to $2\pi \times 37$Hz and $2\pi \times 46$Hz
for $V_0 = 8 E_r$ and $V_0 = 13 E_r$ correspondingly
\cite{Gerbier,Gerbier_private}. We perform simulations in a
sensible range of temperatures. Direct comparison between
experimental and numerical data enables us to estimate the final
temperature of the system. The results of the simulations are sown
in Fig.~\ref{fig:exp}, where the curves for $f_1$,$f_2$, and $f_3$
as functions of the total atom number in the trap $N$ are
parameterized by the temperature.

Thermal effects vanish and a system can be considered to be in its
ground state, if the temperature is substantially smaller than the
energy of the low-lying excitations. In the SF regime, these modes
have typical frequencies of order $\sqrt{m/m_*} \, \omega_{0}$
($m^*$ is the effective mass in the lattice), which for the
parameters of Fig.~\ref{fig:exp}(a)-(c), gives a rough estimate $T
\ll 3t$. Clearly, the curves for $T=t/3$ and $T=t$ in
Fig.~\ref{fig:exp}(a)-(c) are in the regime of effective zero
temperature. Already at $T=2t$ the thermal effects become
significant resulting in a larger atom cloud and, consequently,
reduced average density. At $T < 5t$ the size of the cloud for the
maximum number of atoms in the trap is $\sim 100$ lattice sites in
each dimension. To obtain $f_n$ at $T=10t$, we have to resort to
finite-size scaling, being limited by computer memory at the
linear system size of 150.

In the MI phase, the zero temperature regime is reached for
temperatures smaller than the excitation gap, which is of order
$U/2$ for the parameters of Fig.~\ref{fig:exp}(d)-(f). This leads
to the condition $T \ll 20t$, consistent with the saturation of
$f_n$ below $T=5t$ (see Fig.~\ref{fig:exp}(d)-(f)). For $N
\lesssim 5 \times 10^4$, in the zero temperature limit, the curves
$f_n$ are flat, corresponding to the filling of the $\nu=1$ Mott
shell, and the number distribution is essentially squeezed. Simple
estimates \cite{Gerbier} show that the decrease of $f_1$
accompanied by the increase in $f_2$ at $N \sim 5 \times 10^4$,
and the saturation of $f_2$ with a peak in $f_3$ at $N \sim 2
\times 10^5$ are consistent with the formation of Mott plateaus
with $\nu=2$ and $\nu=3$, respectively. As seen from
Fig.~\ref{fig:exp}(d),(e), final temperature effects degrade the
degree of number squeezing in the MI by favoring particle-hole
excitations.

The comparison of the calculated fraction of atom pairs $f_2$ with
the measurements of Ref.~\cite{Gerbier} (open circles in
Fig.~\ref{fig:exp}(b),(e)) gives the typical experimental
temperatures of order of $5t \approx 1.5\times 10^{-1} E_r$ in the
SF regime and $10t \approx 10^{-1}E_r$ in the MI regime.
Temperatures on the order of a few $t$ have been also observed in
a (one-dimensional) Tonks-Girardeau gas \cite{Bloch_Shlyapnikov},
where the effective fermionization due to strong interactions
allows to deduce the experimental temperature from the momentum
distribution. Note that the system acquires a finite temperature
as a result of its loading into the optical lattice, and therefore
the final temperature is supposed to depend on the number of atoms
in the trap. However, the fact that the experimental data lie
above the $T=0$ curve in Fig.~\ref{fig:exp}(b) at high $N$ is
unlikely to be explained by the the effects of heating. When a
large $f_3$ fraction is present, a change in the spin resonance
condition can result in a considerable contribution of spin
collisions on triply occupied sites to the observed spin
oscillation amplitude \cite{Gerbier}, which could explain the
anomalously high apparent $f_2$ \cite{Gerbier_private}. Such
drifts of the resonance parameters are carefully checked for, but
can not be ruled out completely \cite{Gerbier_private}.

%%%%%%%%%%%%%%%%%%%%%%%%%%%%%%%%%%%
\section{Conclusions}
We studied the ground-state on-site number statistics of
interacting lattice bosons. We considered the limits of weak
interatomic interactions, the limit of large filling in the SF
regime, and the limit of strong interatomic interactions. At $\nu
U/t \ll 1$, the correction to the Poisson distribution is
described by Eq.~(\ref{Pn_relative}), with an essentially
dimension-dependent scaling---Eq.~(\ref{lambda_1D}) in 1D,
Eq.~(\ref{lambda_2D}) in 2D, and Eq.~(\ref{lambda_3D}) in 3D. In
the case of large occupation numbers, $\nu \gg 1$, we show that,
in the region of interactions $U/t \ll \nu$, the on-site
occupation number distribution is Gaussian and its variance, given
by Eq.~(\ref{sigma-zeroT}), gradually decreases with the
asymptotic form $\sigma^2 \propto \sqrt{\nu t/U}$ at $1/\nu \ll
U/t \ll \nu$ in all dimensions. An excellent agreement with
numeric simulations is found already at $\nu=5$. At $\nu t/ U \ll
1$ and integer filling $\nu$, the distribution,
Eq.~(\ref{Pn_Mott}), is dominated by particle-hole fluctuations on
top of an ideal MI. At $\nu=1$, we performed Monte Carlo
simulations of the on-site occupation number distribution in 1D,
2D, and 3D in a wide range of $U/t$ covering the crossover region
between the limiting cases.

We simulated a parabolically confined system in the realistic case
of Ref.~\cite{Gerbier} with the final temperature of the system
being the only free parameter. By direct comparison between
experimental data and numerical results at different temperatures,
we were able to estimate the experimental temperature, which we
found to be of the order of a few $t$ near the transition. The
error bars are small enough to determine $T$ with accuracy of
order $t$. Our results suggest that measurements of the on-site
atom number fluctuations can serve as a reliable method of
thermometry in both superfluid and Mott-insulator regimes. We
believe that with more elaborate techniques, such as that giving
access to the spatial number distribution $P_n(i)$, the
temperature resolution can be further improved. \label{sec:concl}
%%%%%%%%%%%%%%%%%%%%%%%%%%%%%%%%%%%

We are grateful to Fabrice Gerbier for providing us with the
experimental parameters and for fruitful discussions of the
results of this work. This research was supported by the National
Science Foundation under Grant No. PHY-0426881.

\end{document}